\documentclass[aps,prd,showpacs,nobibnotes,twocolumn, 
superscriptaddress,nofootinbib]{revtex4}
\usepackage{graphicx}
\usepackage{amsmath,amssymb}
\usepackage{epstopdf}

\begin{document}

\title{Conservative Constraints on Dark Matter Annihilation into Gamma Rays}

\author{Gregory D. Mack}
\affiliation{Department of Physics, Ohio State University,
Columbus, Ohio 43210}
\affiliation{Center for Cosmology and Astro-Particle Physics,
Ohio State University, Columbus, Ohio 43210}

\author{Thomas D. Jacques}
\affiliation{School of Physics, The University of Melbourne, Victoria 3010, 
Australia}

\author{John F. Beacom}
\affiliation{Department of Physics, Ohio State University,
Columbus, Ohio 43210}
\affiliation{Center for Cosmology and Astro-Particle Physics,
Ohio State University, Columbus, Ohio 43210}
\affiliation{Department of Astronomy, Ohio State University,
Columbus, Ohio 43210}

\author{Nicole F. Bell}
\affiliation{School of Physics, The University of Melbourne, Victoria 3010, 
Australia}

\author{Hasan Y\"{u}ksel}
\affiliation{Department of Physics, Ohio State University,
Columbus, Ohio 43210}
\affiliation{Center for Cosmology and Astro-Particle Physics,
Ohio State University, Columbus, Ohio 43210}

\date{5 May 2008}

\begin{abstract}
Using gamma-ray data from observations of
the Milky Way, Andromeda (M31), and the cosmic
background, we calculate conservative upper limits on the dark matter
self-annihilation cross section to monoenergetic
gamma rays, $\langle \sigma_A v \rangle_{\gamma \gamma}$,
over a wide range of dark matter masses.
(In fact, over most of this range, our results are unchanged if one
considers just the branching ratio to gamma rays with energies
within a factor of a few of the endpoint at the dark matter mass.)
If the final-state branching ratio to gamma rays, $Br(\gamma \gamma)$, were
known, then $\langle \sigma_A v \rangle_{\gamma \gamma} / Br(\gamma \gamma)$
would define an upper limit on the {\it total} cross section; 
we conservatively assume $Br(\gamma \gamma) \gtrsim 10^{-4}$.
An upper limit on the total cross section can also be derived by considering
the appearance rates of {\it any} Standard Model particles; in practice, this limit
is defined by neutrinos, which are the least detectable.
For intermediate dark matter masses, gamma-ray-based and neutrino-based
upper limits on the total cross section are comparable, while the gamma-ray
limit is stronger for small masses and the neutrino limit is stronger for large masses.
We comment on how these results depend on the assumptions about
astrophysical inputs and annihilation final states, and how GLAST and
other gamma-ray experiments can improve upon them.
\end{abstract}



\maketitle


\section{Introduction}

While there is ample gravitational evidence for dark matter, the
nature of these particles remains mysterious and is defined principally
by the weakness of their interactions; for reviews, see e.g.,
Refs.~\cite{Jungman:1995df, Bertone:2004pz,Bergstrom:2000pn}.
The dark matter self-annihilation cross section is of particular importance,
since dark matter concentrations will produce gamma rays and other
detectable Standard Model (SM) particles.  If the dark matter (DM)
is a thermal relic of the early universe, the annihilation cross
section must be $\langle \sigma_A v \rangle \sim 3 \times 10^{-26}$
cm$^3$ s$^{-1}$ in order to obtain the observed relic abundance,
$\Omega_{\rm DM} \simeq 0.3$.  (Throughout, we consider this cross
section averaged with velocity over the dark matter velocity distribution; in
the Milky Way, $v_{rms} \sim 10^{-3}c$.)  It is possible that dark
matter is not a thermal relic, e.g., Ref.~\cite{Kaplinghat:2000vt,Das:2006ht},
which makes it even more
interesting to consider direct late-universe constraints on the
annihilation cross section,
e.g., Refs.~\cite{Jungman:1995df, Bertone:2004pz, Bergstrom:2000pn, 
Ullio:2002pj, Beacom:2006tt, Yuksel:2007ac, Kachelriess:2007aj,
Hooper:2008zg, Bell:2008ey, Dent:2008qy}.

Even if the {\it total} annihilation cross section is set by the
relic abundance, the branching ratios to specific final states are
model-dependent.  The dark matter disappearance rate due to annihilation can be constrained
by the appearance rates of various SM particles.
If the dark matter is the lightest stable
particle in some new physics sector, then it can be natural to have
the final-state branching ratio to SM particles, $Br\textrm{(SM)}$,
be $100\%$, as we assume.  
If the final states include new and purely sterile particles, then all
appearance-based results are weakened proportionally
to $Br\textrm{(SM)}$.

We assume that annihilation is not prevented in principle by dark
matter not being its own antiparticle, or in practice by a large
particle-antiparticle asymmetry.
We also assume that a single type of new particle comprises the dark
matter that is required to exist in the present-day universe, and that,
consistent with observations, the density distributions of dark matter
halos are not appreciably affected by possible dark matter interactions.
These assumptions are made implicitly in nearly all papers about dark
matter annihilation.

Here we calculate the constraints that can be placed on the annihilation
cross section using gamma rays, the most detectable final states, over 
a wide range of dark matter masses.
We first focus on the $\gamma \gamma$ final state, as it would be a
very clean signature of dark matter annihilation, with $E_\gamma = m_\chi$,
e.g., Refs.~\cite{Rudaz:1990rt, Bouquet:1989sr, Pullen:2006sy, Serpico:2008ga,
Profumo:2008yg, Cirelli:2008id, Pospelov:2008qx}.
Unfortunately,
in typical models, this is small, $Br(\gamma \gamma) \sim
10^{-4}-10^{-3}$; in some models, it can be larger
\cite{Matsumoto:2005ui, Gustafsson:2007pc, Ferrer:2006hy}, but one
cannot be certain that these predictions match with nature.
Since gamma rays will be ubiquitously produced, directly in SM final states,
or through radiative corrections and energy-loss processes, we also consider
more general outcomes, in which the gamma-ray energies are in a
broader range below $m_\chi$.

We consider constraints on the dark matter annihilation cross section over a
large mass range of $10^{-5}$ -- $10^5$ GeV.   At all but the highest
energies, gamma-ray data is available to test the 
annihilation cross section, provided that we combine
constraints defined using the Milky Way halo, the Andromeda
halo, and all the halos in the universe.  (Modern data, especially those from
observations of the Milky Way and Andromeda, are significantly more
constraining than those that were available earlier, e.g., at the time of
Ref.~\cite{Ressell:1989rz}, which considered limits on the decay of an
unstable massive neutrino.)  We hope that our results
will be useful in challenging experiments to report stronger limits
using new data and focused analyses.  With the launch of GLAST 
this year, and with new studies by TeV-range experiments,
these prospects are good.  Using our upper limits on
the dark matter annihilation cross section to gamma rays, and a conservative
assumption about the branching ratio to monoenergetic gamma
rays, we define upper limits on the total cross section and
compare to other constraints.

Since the dark matter annihilation rate scales with density squared and the
density profiles are uncertain, we are mindful of how our
constraints on the cross section are affected by astrophysical
uncertainties.  We are conservative in our input choices and
analysis methods, and we show how our
results depend on these.  In light of these considerations, we
do not consider corrections that would affect the results by less than
a factor of $\sim 2$, which also allows some 
simplifications.   Our upper
bounds on the annihilation cross section to gamma rays
would only be improved by more optimistic assumptions.

In Section II, we discuss important general bounds on the total
annihilation cross section.  In Section
III, we review the analysis methods used for the case of gamma-ray 
lines from various dark matter concentrations.  The 
experiments and observations we use are discussed in Section IV.
In Section V, we summarize and interpret our results.


\section{Cross Section Constraints}

The annihilation cross section sets the dark matter
{\it disappearance} rate, for which there are two important constraints.  The
first is unitarity~\cite{Griest:1989wd, Hui:2001wy}, which sets a
general upper bound that can only be
evaded in unusual cases~\cite{Kusenko:2001vu}.  In the low-velocity limit,
where s-wave annihilation dominates, the unitarity bound is $\langle
\sigma_A v \rangle < 4 \pi/m_\chi^2 v$, or
\begin{equation}
\langle \sigma_A v \rangle \leq
\left(1.5 \times 10^{-13} ~\textrm{cm}^3 ~\textrm{s}^{-1}\right)
\left[\frac{300 ~\textrm{km/s}}{v_{rms}}\right]
\left[\frac{\textrm{GeV}}{m_\chi}\right]^2\,.
\end{equation}
For $m_\chi \agt 10^6$ GeV, this would require that
$\langle \sigma_A v \rangle$ be smaller than that for a thermal relic,
and which would thus give too large of a relic abundance~\cite{Griest:1989wd}.
However, for smaller $m_\chi$ the unitarity limit is much less constraining.
The second general constraint comes from the requirement that annihilation
does not drastically alter the density profiles of dark matter halos in the
Universe today.  In the model of Ref.~\cite{Kaplinghat:2000vt}, a
large self-annihilation cross section was invoked in order to
reconcile predicted cuspy density profiles with the flatter
ones inferred from observation, requiring
\begin{equation}
\langle \sigma_A v \rangle_{\textrm{KKT}} \simeq
\left(3 \times 10^{-19} ~\textrm{cm}^3 ~\textrm{s}^{-1}\right)
\left[\frac{m_\chi}{\textrm{GeV}}\right]\,.
\end{equation}
We re-interpret this result as an approximate
upper bound, beyond which halo density
profiles would be significantly distorted by dark matter annihilation.  Note
that this limit is very weak for all but the lightest masses.

We now discuss limits which arise from the {\it appearance} rate of dark matter
annihilation products, assuming $Br\textrm{(SM)} = 100\%$. 
All final states except neutrinos obviously produce gamma
rays, either directly or as secondary particles (we return to neutrinos next).
Quarks and gluons
hadronize, producing pions and thus photons via $\pi^0 \rightarrow
\gamma\gamma$, while $\tau^\pm$, $W^\pm$ and $Z^0$ also produce $\pi^0$
via their decays.
Charged particles produce photons via
electromagnetic radiative corrections~\cite{RC}, and electrons and
positrons also produce photons via energy loss processes~\cite{Eloss}.
Therefore we expect a broad spectrum of final state
photons, even though the branching ratio to the monoenergetic
$\gamma\gamma$ final state may be small.
We use the gamma-ray data to place upper limits on
$\langle \sigma_A v \rangle_{\gamma \gamma}$,
and these are of general interest for their own sake.  With an assumption
about $Br(\gamma \gamma)$, these results also define an upper limit
on the total cross section of
$\langle \sigma_A v \rangle_{\gamma \gamma} / Br(\gamma \gamma)$.
Unless one is confident that $Br(\gamma \gamma)$ cannot be too small,
then this constraint on the total cross section can be arbitrarily weakened.

An important general limit on the {\it total} annihilation cross
section can be obtained by considering annihilation into the {\it least}
detectable final state, namely neutrinos~\cite{Beacom:2006tt, Yuksel:2007ac}.
Given that stronger constraints will exist on all final states other than
neutrinos, we can set a conservative upper bound on the total dark matter
annihilation rate by assuming the branching ratio to neutrinos is
100\%.  (Unlike all other constraints, the neutrino constraint, being
the weakest, is not to be divided by a realistic branching ratio; this
follows from the fact that the sum of all branching ratios must be
100\%.)   The resulting limits are surprisingly strong.
Dark matter annihilation into neutrinos was explored by
Beacom, Bell, and Mack (BBM)~\cite{Beacom:2006tt}, and Y\"{u}ksel
et al.~(YHBA)~\cite{Yuksel:2007ac}, for cosmic and Galactic dark matter sources,
respectively.  By requiring that the neutrino flux produced by
annihilation be smaller than the measured atmospheric neutrino
background, robust bounds on the total annihilation cross section were
obtained over a wide mass range.  (Ref.~\cite{PalomaresRuiz:2007eu}
extended the dark matter annihilation limits to lower masses and
Ref.~\cite{PalomaresRuiz:2007ry} developed analogous dark matter
decay limits over a wide range of masses.)
For all masses considered, these
limits are much stronger than the KKT limit; they are also stronger
than the unitarity limit except at high masses.
While neutrinos are the least detectable annihilation products, even
they are accompanied by gamma rays via electroweak radiative
corrections; these results lead to constraints on
$\langle \sigma_A v \rangle$ that are comparable to or better than
those obtained directly with neutrinos~\cite{Kachelriess:2007aj,
Bell:2008ey, Dent:2008qy}.


\section{Calculation of Dark Matter Signals}

The dark matter annihilation rate depends on the square of the dark matter
number density $\rho/m_\chi$, which is written in terms of the unknown
mass $m_\chi$ and the uncertain dark matter mass density $\rho$.  Not
coincidentally, where the density is largest, at the centers of halos,
the uncertainties are the largest; these regions contribute relatively
little to the gravitationally-measured mass of a halo.  To cover as
large of an energy range as possible, we have to consider gamma-ray
data for the Milky Way, Andromeda, and all of the dark matter halos in the
universe.  In all cases, though the astrophysical and analysis
uncertainties vary in their severity, we make conservative choices for
the dark matter density and hence the cross section limits (smaller choices
for the density mean larger upper limits on the cross section).


\subsection{Dark Matter Halos}

A standard parameterization of the dark matter density profile in a halo is
\begin{equation}
\rho(r) =
\frac{\rho_0}{\left(r/r_s\right)^{\gamma}
\left[1+\left(r/r_s\right)^\alpha\right]^{(\beta - \gamma)/\alpha}}\,.
\end{equation}
The Navarro-Frenk-White (NFW)~\cite{NFW} and Kravtsov
profiles~\cite{Kravtsov:1997dp} are defined by $(\alpha, \beta,
\gamma) = (1, 3, 1)$ and $(\alpha, \beta, \gamma) = (2, 3, 0.4)$
respectively.  Near the center of the halo, the density of an NFW
profile scales with radius as $1/r$, while the Kravtsov profile scales
less steeply, as $1/r^{0.4}$.  For large radii, $r \agt r_s$, these
two profiles coincide more closely.  For the Milky Way, $r_s$ is 20
kpc for the NFW profile and 10 kpc for the Kravtsov profile, while the
normalization, $\rho_0$, is fixed such that the density at the solar
circle distance, $R_{sc}=8.5$ kpc,
is $\rho(R_{sc} ) = 0.3$ GeV cm$^{-3}$ (0.37 GeV cm$^{-3}$) for the
NFW (Kravtsov) profile.

There does not appear to be a consensus on the values of the halo
parameters for Andromeda; for example, compare the NFW profiles in
Ref.~\cite{Fornasa:2007ap} with Ref.~\cite{Fornengo:2004kj}, where
both $\rho_0$ and $r_s$ are quite different. Thus we have
chosen to model Andromeda using the Milky Way parameters, as an
appropriate compromise between competing extremes.

In the innermost regions of halos, the uncertainties in the dark matter
density and thus the
annihilation rate are at their largest.  For larger central regions, or whole
halos, these uncertainties are much less.  For the Milky Way, these
effects can easily be seen in Fig.~2 of YHBA; for larger angular
regions centered on the Galactic Center, the dark matter annihilation signals
for different profiles become much more similar to each other.  While
the uncertainties at small angular scales can be orders of magnitude,
those at large angular scales are not more than a factor of about 2.

In YHBA, we explored in detail how various annihilation signals
depend on the choice of dark matter density profile.   Our overall
approach is to be conservative by adopting smaller choices of
the astrophysical inputs; this means that larger values of the cross
section would be required to get the same gamma ray or neutrino
fluxes.  Here we use the Kravtsov profile for our main results; for
the commonly-adopted NFW profile, we find smaller (more restrictive)
upper bounds on the cross sections, as shown below.
Also to be conservative, we neglect the possibility of halo substructure,
e.g., Refs.~\cite{Diemand:2006ik, Strigari:2007at},
or mini-spikes around intermediate-mass
black holes~\cite{Bertone:2005xz, Horiuchi:2006de}, which would lead to
enhanced annihilation signals.


\subsection{Milky Way and Andromeda Signals}

We first consider annihilations in our Galaxy, following the
conventions of YHBA, and generalize this to the nearby galaxy
Andromeda (M31).  The intensity (flux per solid angle) of the annihilation
signal at an angle $\psi$ with respect to the Galactic Center (GC) is
proportional to the square of the dark matter density integrated over the line
of sight,
\begin{equation}
{\cal J}(\psi) = {\rm J_0}
\int^{\ell_{max}}_0 \rho^2\left(\sqrt{R_{\textrm{sc}}^2 -
2\ell R_{\textrm{sc}}\cos{\psi} +\ell^2}\right)d\ell \,,
\label{los}
\end{equation}
where ${\rm J_0}=1/[ 8.5 \, {\rm kpc} \times (0.3 \, {\rm GeV \, 
cm}^{-3})^2]$ is an arbitrary normalization we use to make ${\cal J}$ a
dimensionless quantity, and which cancels in the final results.
The upper limit of the integration is given by
$\ell_{max} = (R^2_{\textrm{MW}} - \sin^2{\psi}R_{\textrm{sc}}^2)^{1/2}
+ R_{\textrm{sc}}\cos{\psi}$.
We define ${\cal J}_{\Delta\Omega}$ as the average of ${\cal J}$
over a cone of half-angle $\psi$ centered on the GC,
\begin{equation}
{\cal J}_{\Delta \Omega} = \frac{2\pi }{\Delta \Omega} \int_0^{\psi}
{\cal J}(\psi )\sin{\psi }d\psi\,,
\end{equation}
where $\Delta\Omega = 2\pi(1-\cos{\psi})$ is the angular size
of the cone in steradians.  The values of ${\cal J}(\psi)$ and
${\cal J}_{\Delta\Omega}$ can be read directly from Fig.~2 of YHBA
(below, we do not explicitly show the sr$^{-1}$ units of
 ${\cal J}(\psi)$  and ${\cal J}_{\Delta\Omega}$).

Eq.~(\ref{los}) can easily be generalized to external
halos~\cite{Falvard:2002ny,Evans:2003sc} (such as the Andromeda galaxy
at a distance of $D_{\textrm{M31}} \simeq 700$ Mpc) using
\begin{equation}
{\cal J}(\psi) =  {\rm J_0}
\int^{\ell_{max}}_{\ell_{min}}
\rho^2\left(\sqrt{D_{\textrm{M31}}^2 -2\ell D_{\textrm{M31}}\cos{\psi} +\ell^2}\right) d\ell \,,
\label{los2}
\end{equation}
where the result is independent of the upper and lower limits of
integration ($\ell_{min}, \ell_{max}$) as long as they cover most of the
halo under consideration.

For extragalactic dark matter sources, the annihilation signals will include a
contribution from the dark matter in our own galaxy along the line of
sight.  However, in the case of an external galaxy like Andromeda, this
contribution will be eliminated if there is a subtraction of the background
intensity from a region close to the source, as is often done in observational
analyses.

With these definitions, the intensity
of the dark matter annihilation gamma-ray signal is
\begin{equation}
\frac{d\Phi_{\gamma}}{dE} =
\frac{\langle \sigma_A v\rangle}{2}
\frac{{\cal J}_{\Delta \Omega}}{{\rm J_0}}
\frac{ 1 }{4\pi m_\chi^2}
\frac{dN_{\gamma}}{dE}\,,
\label{haloflux}
\end{equation}
where $dN_\gamma/dE$ is the gamma-ray spectrum per annihilation.  In the
case of annihilation into two monoenergetic gamma rays, we simply have
$dN_{\gamma}/dE = 2\delta(m_\chi - E)$; we generalize this below.
Similarly, the total flux (per unit energy) from a region of solid
angle $\Delta\Omega$ is
\begin{equation}
\frac{d\phi_{\gamma}}{dE} =
\frac{d\Phi_{\gamma}}{dE} \Delta\Omega  =
\frac{\langle \sigma_A v\rangle}{2}
\frac{{\cal J}_{\Delta \Omega} \Delta\Omega}{{\rm J_0}} 
\frac{ 1 }{4\pi m_\chi^2}
\frac{dN_{\gamma}}{dE} \,.
\label{haloflux2}
\end{equation}


\subsection{Cosmic Diffuse Signal}
\label{cosmic}

The calculation of the cosmic diffuse annihilation signal is detailed,
for example, in Refs.~\cite{Bergstrom:2001jj,Ullio:2002pj}, where the
cosmological flux of annihilation products from external galaxies was
calculated taking the clustering of dark matter into account.
Recently, BBM and YHBA applied this technique to the scenario where
dark matter annihilates into neutrinos.

The cosmic diffuse flux, arising from dark matter annihilation in halos
throughout the Universe, is
\begin{eqnarray}
\frac{d\Phi_\gamma}{dE} &=& \frac{\langle \sigma_A v \rangle}{2}
\frac{c}{4 \pi H_0} \frac{\Omega^2_{DM}
\rho^2_{\textrm{crit}}}{m_\chi^2}\nonumber \\
&\times& {\int_0}^{z_{up}} \frac{f(z)(1+z)^3}{h(z)}
\frac{dN_\gamma(E')}{dE'} e^{-z/z_{max}}dz\,,
\label{cosmicflux}
\end{eqnarray}
where $H_0 = 70$ km s$^{-1}$ Mpc$^{-1}$ is the Hubble parameter and
$\Omega_{DM}$ is the dark matter density in units of the critical
density.  We assume a flat universe, with $\Omega_{DM}$ = 0.3,
$\Omega_\Lambda$ = 0.7, $h(z) = [(1+z)^3 \Omega_{DM} +
\Omega_\Lambda]^{1/2}$.  The factor $e^{-z/z_{max}}$, taken from
Ref.~\cite{Ullio:2002pj}, accounts for the attenuation of gamma rays,
a modest effect for the energies considered here.  
The factor $f(z)$ in Eq.~(\ref{cosmicflux}) accounts for the average
increase in density squared due to the fact that dark matter is clustered into
halos, rather than uniformly distributed, and the evolution with
redshift of the halo number density.  (The $\Delta^2$ factor in BBM is
equal to $f(z)(1+z)^3$.)  Following YHBA, we use the parameterization
$\log_{10}(f(z)/f_0) = 0.9 \,[\textrm{exp}(-0.9z)-1]-0.16z$, where $f_0$
depends on the halo profile.  Choosing the Kravtsov (NFW) profile,
$f_0 \simeq 2 \, (5) \,\times 10^{4}$.

Gamma rays that are produced with energy $E'$ are observed with
redshifted energy $E = E^\prime /(1+z)$.  For annihilation into
monoenergetic gamma rays, the delta function source spectrum is modified
by redshift as
\begin{equation}
\frac{dN}{dE'} = 2~\delta(m_\chi - E^\prime) =
\frac{2}{E}~\delta\left(z - (\frac{m_\chi}{E} - 1)\right)\,,
\end{equation}
which shows that the observed flux at an energy $E$ is
contributed by sources at redshift $\frac{m_\chi}{E} - 1$.


\section{Specific Observations and Derived Annihilation Constraints}

We have collected gamma-ray flux measurements and limits from a wide
variety of experiments, spanning an extensive energy range from 20 keV
to 10 TeV. In most of the observations, the energy spectra are given
in log-spaced energy intervals.  We calculate annihilation gamma-ray
fluxes for the Galactic, Andromeda, and cosmic dark matter sources, using the methods
outlined in Section III above.  These are compared with observational
data over an energy range, conveniently chosen as
$10^{-0.4}m_\chi$ -- $m_\chi $, that is comparable to or larger than the
energy resolution and bin size of the experiments.  If only upper
limits on the flux are given, we instead compare our predictions
directly with these upper limits.

\begin{figure}
\includegraphics[width=3.25in, clip=true]{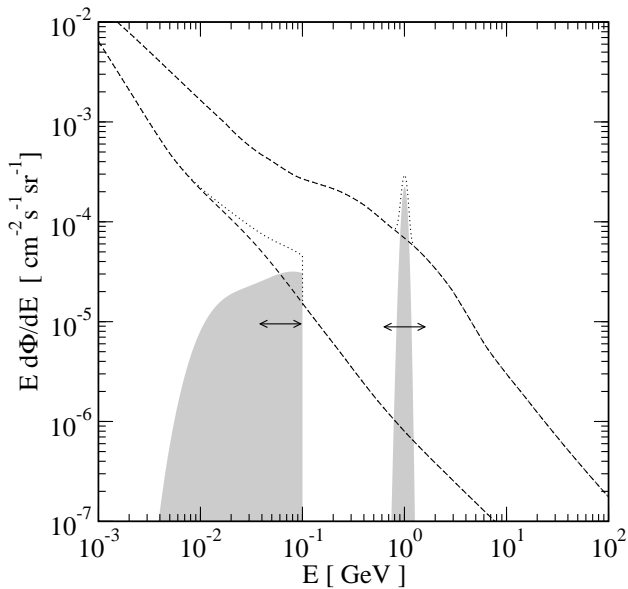}
\caption{Example dark matter annihilation signals,
shown superimposed on the Galactic
and extragalactic gamma-ray spectra measured by COMPTEL and EGRET.
In each case, the cross section is chosen so that the signals are 
normalized according to our conservative detection criteria, namely,
that the signal be 100\% of the size of the background when integrated in the
energy range chosen (0.4 in $\log_{10}E$, shown by horizontal arrows).
The narrow signal on the right is the Galactic Center flux due to
annihilation into monoenergetic gamma rays, for $m_\chi = 1$ GeV;
the signal is smeared as appropriate for a detection with finite
energy resolution.  The broad feature on the left is the cosmic
diffuse signal for annihilation into monoenergetic gamma rays at
$m_\chi = 0.1$ GeV, smeared by redshift.}
\label{fig:sigmagg-fig1}
\end{figure}

Our constraints on the dark matter annihilation rate are conservatively determined
by demanding that the annihilation flux be smaller than 100\% of the
observed (presumably not produced by dark matter) gamma-ray background flux at the
corresponding energy range.  In Fig.~\ref{fig:sigmagg-fig1}, we show the GC
and cosmic diffuse signals from dark matter annihilations which fulfill this
criterion, superimposed upon the Galactic and extragalactic spectra,
respectively, as measured by COMPTEL and EGRET.

The experiments report their results as either intensity (as in
Eq.~\ref{haloflux}), which requires that we calculate ${\cal J}_{\Delta
\Omega}$, or flux from a given angular region (as in
Eq.~\ref{haloflux2}), for which we need ${\cal J}_{\Delta \Omega}
\Delta \Omega$.  We present the values of these parameters which
correspond to the Kravtsov profile, as this results in the most
conservative upper limits on the annihilation cross section. 
Our limits on the dark matter
annihilation cross section are reported in Fig.~\ref{fig:sigmagg-fig2},
where we also show how our results would change if the NFW profile
were adopted instead.  The
details of the experiments and our analyses are summarized below for
each observation.



\subsection{COMPTEL and EGRET}
\label{COMPEGRET}

COMPTEL~\cite{COMPTELweb}, the imaging COMPton TELescope aboard the
Compton Gamma Ray Observatory (CGRO) satellite, measured gamma rays in
the energy range 1--30 MeV.  EGRET~\cite{EGRETweb}, the Energetic
Gamma Ray Experiment Telescope, also aboard the CGRO, measured gamma
rays in the energy range 30 MeV to nearly 100 GeV.  For both COMPTEL
and EGRET, the full sky was studied with an angular resolution of at
worst a few degrees (for the large regions we consider, this makes no
difference).  The energy resolution was modest, and the data were
given in a few logarithmically-spaced bins per decade in energy.

Both COMPTEL and EGRET observed the Galactic Center region, and the
measured gamma-ray intensity energy spectra are reported in
Ref.~\cite{Strong:1998ck, Strong:2005zx}
for the region $-30^\circ < l < 30^\circ$
and $-5^\circ < b < 5^\circ$ (Galactic longitude and latitude,
respectively).  The disk-like morphology of the emission region makes
it clear that nearly all of this emission is due to ordinary
astrophysical sources; to be conservative, we do not attempt to
define a limit on the small component of this that could be due to
centrally-concentrated dark matter, and simply use the total observed
intensity to bound any dark matter contribution.  Also, we
evaluate the dark matter signal as if from a circular region of $\psi
= 30^\circ$; accounting for the rectangular shape of the region would
lead to a higher value than the ${\cal J}_{\Delta\Omega} \simeq 13$ that we
adopt.  Using a less conservative set of assumptions than we employ, stronger
limits on $\langle \sigma v\rangle_{\gamma\gamma}$ were derived from
the EGRET data in Ref.~\cite{Pullen:2006sy}.


\subsection{H.E.S.S.}

H.E.S.S. (High Energy Stereoscopic System), a system of multiple
atmospheric \v{C}erenkov telescopes, is presently in operation in
Namibia~\cite{HESSweb}.
H.E.S.S. observed the Galactic
Center region in the energy range 0.3--15 TeV.  An apparent point
source at the Galactic Center was observed, as was an extended
source ($\sim 1^\circ$) known as the Galactic Center
Ridge~\cite{Aharonian:2006au}.  While the origin of the point
source is unknown, the Ridge emission is almost certainly
astrophysical, and is consistent with being caused by cosmic rays
colliding with a gas cloud (again, we do not attempt to account for
this, and will simply bound any dark matter contribution by the total
observed intensity).

Since the uncertainties in the dark matter profile increase for
smaller angular regions around the Galactic Center, it is more robust
to define our results using the extended region instead of the point
source.  The Ridge emission was observed in an angular region
$-0.8^\circ < l < 0.8^\circ$ and $-0.3^\circ < b < 0.3^\circ$ in
Galactic coordinates, and the resulting flux reported by H.E.S.S.
reflected a background subtraction from a nearby region ($-0.8^\circ <
l < 0.8^\circ$ and $0.8^\circ < b < 1.5^\circ$) to help account for
cosmic rays.  Thus, we have to consider not the whole dark matter
signal, but just its contrast between the central and adjacent regions
by accounting for this subtraction in our analysis.

We approximate the intensity from the
rectangular region of the Galactic Center Ridge with
a circle of radius $0.8^\circ$.
We also estimate
the adverse effect of the background subtraction on our limits
by choosing $ {\cal J}$ to be subtracted at its maximum, 
i.e., $\psi =0.8^\circ$. This means
\begin{equation}
{\cal J}_{\Delta\Omega} =
\frac{2\pi}{\Delta\Omega}
\int_0^{0.8^\circ} ({\cal J}(\psi) -
{\cal J}(0.8^\circ))\sin{\psi}d{\psi} \simeq 3\,.
\end{equation}
Had we not made this subtraction correction, our limits on the
cross section would be stronger by about an order of magnitude.


\subsection{INTEGRAL}

The space-borne INTEGRAL (INTErnational Gamma-Ray Astrophysics
Laboratory) observatory~\cite{INTweb} has searched for gamma-ray
emission in the Milky Way over the energy range 20--8000 keV, using
the SPectrometer on INTEGRAL (SPI).  Teegarden and
Watanabe~\cite{Teegarden:2006ni} presented results of an INTEGRAL
search for gamma-ray line emission from the Galactic Center region (we
use their zero-intrinsic-width results, as appropriate for the low
dark matter velocities of the halo).
Other than the expected positron
annihilation~\cite{Beacom-Yuksel-positrons} and $^{26}$Al
decay~\cite{Diehl:2006cf}
signatures, no evidence of other line emission was found.

To reduce backgrounds and improve the sensitivity of the line search,
the measured intensity from large angular radii ( $> 30^{\circ}$) was
subtracted from that in the Galactic Center region ( $< 13^{\circ}$),
resulting in a 3.5-$\sigma$ constraint on the flux of very roughly
$\lesssim$ 10$^{-4}$ photons cm$^{-2}$ s$^{-1}$ in the energy range
20--8000 keV.  Our calculations must reflect this subtraction, which
will somewhat weaken the sensitivity to the dark matter signal.  A
similar correction was used in Ref.~\cite{Yuksel:2007xh}.  We
implement this as
\begin{equation}
{\cal J}_{\Delta\Omega} \Delta\Omega =
2\pi \int_0^{13^\circ}({\cal J}(\psi) -
{\cal J}(>30^\circ))\sin{\psi} d\psi \simeq 2\,.
\end{equation}
Due to the decreasing trend of the dark matter profile, the intensity
outside the Galactic Center region will be largest at $30^\circ$, and
accordingly we choose this value to be as conservative as possible (a
larger subtraction leads to a weaker upper limit on $\langle \sigma_A
v \rangle$).  Had we not made this correction, our limits on the
cross section would be stronger by about a factor of 2.


\subsection{Andromeda Halo Results}

The Andromeda galaxy (M31) has been observed by several gamma-ray
experiments, all of which placed upper limits on the flux.
EGRET, CELESTE, and HEGRA all observed Andromeda, each encompassing a
respectively smaller angular region of that extended object.
As the results were reported as flux limits from specified angular regions,
we compare to these using ${\cal J}_{\Delta\Omega} \Delta\Omega$,
which is an input to Eq.~(\ref{haloflux2}).

EGRET viewed Andromeda with an angular radius of $0.5^\circ$ and set a
2-$\sigma$ upper limit on the gamma-ray flux of $1.6 \times 10^{-8}$
photons cm$^{-2}$ s$^{-1}$ from 0.1 GeV to 2 GeV, since no signal was
seen~\cite{Blom:1998iu}. For the angular region of this observation,
the flux will be proportional to
\begin{equation}
{\cal J}_{\Delta\Omega} \Delta\Omega =
2\pi \int_0^{0.5^\circ}{\cal J}^\prime (\psi)\sin{\psi} d\psi \simeq 2 \times 10^{-3}\,.
\end{equation}

CELESTE (\v{C}Erenkov Low Energy Sampling and Timing Experiment) is an
atmospheric \v{C}erenkov telescope in the French Pyrenees, which
studies gamma rays with energies greater than 50
GeV~\cite{CELESTEweb}.
It viewed Andromeda in the energy range of 50--700 GeV, and again no signal
was seen~\cite{Lavalle:2006rs}.  A 2-$\sigma$ upper limit on the
energy-integrated flux from Andromeda was reported as $\lesssim 10^{-10}$
photons cm$^{-2}$ s$^{-1}$; employing an angular radius of
$\theta_{\textrm{obs}} = 0.29^\circ$ yields ${\cal J}_{\Delta\Omega}
\Delta\Omega \simeq 1 \times 10^{-3}$.

HEGRA (High Energy Gamma Ray Astronomy experiment) was an atmospheric
\v{C}erenkov telescope, located in La Palma in the Canary
Islands~\cite{HEGRAweb}.  It took data in the range 0.5--10 TeV, with
better energy resolution than that of CELESTE~\cite{Aharonian:2003xh}.
It used an even smaller angular radius of
$\theta_{\textrm{obs}} = 0.105^\circ$, which yields $ {\cal
J}_{\Delta\Omega} \Delta\Omega \simeq 2 \times 10^{-4} $.
HEGRA reported 99\% C.L.~upper limits for the gamma-ray line flux,
and these can be used directly.


\subsection{Cosmic Diffuse Results}

INTEGRAL~\cite{integralcosmicdiffise},
COMPTEL~\cite{comptelcosmicdiffuse} and EGRET~\cite{Strong:2004ry}
have all made measurements of the gamma-ray flux at high latitudes,
and these can be used to set a limit on the cosmic dark matter
annihilation signal.   The INTEGRAL data used here were those collected
in broad energy bins, much like those of COMPTEL and EGRET.
The cosmic gamma-ray background was also measured by the Gamma-Ray
Spectrometer aboard the Solar Maximum Mission (SMM)
\cite{SMMweb} over the energy range 0.3 -- 8 MeV, for a field of view
of 135$^\circ$ in the direction of the Sun~\cite{SMM}, and we include this
data.

For the cosmic diffuse analysis, the framework detailed in
Section~\ref{cosmic} can be applied.  
Note that for simplicity we calculate only the true cosmic diffuse dark matter
signal, neglecting any Galactic contribution along the lines of sight.
This contribution from the Galactic halo (which would add to the signal
and thus make our limits stronger) is significant for NFW or steeper
profiles and can even dominate over the true cosmic dark matter signal; see
YHBA and Ref.~\cite{Yuksel:2007dr}.


\section{Discussion and Conclusions}


\subsection{Limits on the Cross Section to Gamma Rays}

\begin{figure}
\includegraphics[width=3.25in, clip=true]{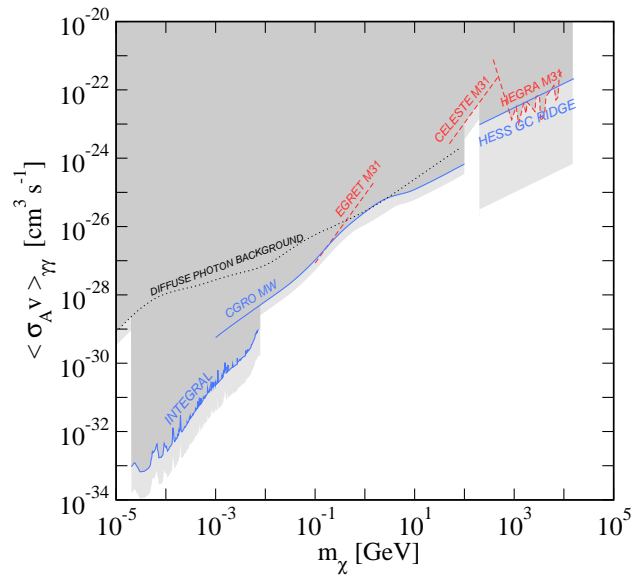}
\caption{
  The limits on the partial cross section, $\langle \sigma_A v
  \rangle_{\gamma\gamma}$, derived from the various gamma-ray data.
  Our overall limit is shown as the dark shaded exclusion region.  For
  comparison, the light-shaded region shows the corresponding limits
  for the NFW (rather than the Kravtsov) profile.}
\label{fig:sigmagg-fig2}
\end{figure}

In Fig.~\ref{fig:sigmagg-fig2}, we combine all of the upper limits on the
partial cross section to monoenergetic gamma rays, choosing the
strongest limit for each value of the dark matter mass.  The shaded
exclusion region shows our combined bound.  These searches for
dark matter signals are limited by astrophysical backgrounds, and the
general trend of how the limits vary with mass follows from how these
backgrounds vary with energy.
We can estimate how the cross section limit should scale with mass,
and how it should depend on the assumed spectrum of final-state
gamma rays and the choice of density profile.

Recall that we conservatively require the signal to be as large as the
full measured background in an energy bin. 
The  gamma-ray number flux of the signal integrated in a logarithmic
energy bin $\Delta(\ln E)$ scales as
$E d\Phi/dE \, \Delta(\ln E) \sim \langle \sigma_A v \rangle / m_\chi^2$,
provided that the bin is wide enough to contain the full signal.
The gamma-ray number flux of the background integrated in the same
logarithmic energy bin scales as
$E d\Phi/dE \, \Delta(\ln E) \sim E/E^\alpha \, \Delta(\ln E)$, for a background
spectrum $d\Phi/dE \sim 1/E^\alpha$.
For a narrow bin, the evaluation point is $E \sim m_\chi$.  We then expect
the upper limit on the cross section to scale as
$\langle \sigma_A v \rangle_{limit} \sim m_\chi^{3-\alpha} \, \Delta(\ln E)$.
For example, for the EGRET diffuse data,
$\alpha$ is slightly greater than 2, and so the cross section limits
in this energy range scale slightly less rapidly than as $\langle \sigma_A v
\rangle_{limit} \sim m_\chi$.

Most of these experiments had modest energy resolution.  To
be conservative, we assume an analysis bin with a logarithmic
energy width of 0.4 in $\log_{10} E$ (i.e., $\Delta(\ln E) \sim 1$)
for the Galactic and cosmic
diffuse analyses; this is at least as wide as the energy bins
reported by the experiments.  That is, even though we nominally
assume two monoenergetic gamma rays at $E_\gamma = m_\chi$,
our results have not taken advantage of this fact.  In effect, our
results are what one would obtain for an annihilation gamma-ray
spectrum as  wide as 0.4 in $\log_{10} E$.  The exception is the
INTEGRAL line search, where the excellent energy resolution is what
leads to this limit being stronger than expected from the general
trend in Fig.~\ref{fig:sigmagg-fig2}.

Due to radiative corrections~\cite{RC} or energy-loss
processes~\cite{Eloss}, there should be some gamma rays
near the endpoint, and our results can be scaled if the
assumed branching ratio is less than the 100\% used in
Fig.~\ref{fig:sigmagg-fig2}.   For example, for charged-particle final
states, the branching ratio to internal bremsstrahlung gamma rays
near the endpoint is $Br(\gamma) \sim \alpha \sim 10^{-2}$.  For
neutral final states, there will typically be gamma rays (or neutrinos)
near the endpoint.  To be conservative about these details, we chose
a nominal minimum branching ratio to gamma rays near the endpoint
of $10^{-4}$.

How would our results change if we considered an even broader
annihilation gamma-ray spectrum?
We emphasize that the results shown in Fig.~\ref{fig:sigmagg-fig2},
which are based on direct numerical integration, are already valid
for spectra as wide as our analysis bins.  First, we should take into
account the increase in the logarithmic bin width.  Second, to
be more precise, the evaluation point for the background spectrum
should not be $E = m_\chi$, but rather $E = m_\chi/a$, with $a > 1$.
This increases the estimate of the integrated background, and hence
the cross section limit, by a factor $\sim a^{\alpha - 1}$.  Thus, if we took
the annihilation gamma-ray
spectrum to be as much as one order of magnitude wide, then
our limits in Fig.~\ref{fig:sigmagg-fig2} would be weakened by at
most a factor of several, depending on the background spectrum.
(For the INTEGRAL line search, the correction would be much larger.)

Given the large range on the axes in
Fig.~\ref{fig:sigmagg-fig2}, and our intention to define approximate and
conservative limits, this shows that our results are much more general
than they first appear.  Similarly, the results in BBM~\cite{Beacom:2006tt}
and YHBA~\cite{Yuksel:2007ac} do not have a strong dependence on
assumed annihilation neutrino spectrum.

How sensitive are our limits to the choice of density profile?  As
noted, we chose the rather shallow Kravtsov profile to be
conservative.  If we were to adopt an NFW profile, which increases much more
rapidly toward the Galactic Center (scaling with radius as $r^{-1}$
rather than $r^{-0.4}$) the annihilation rates would be larger and the
cross section limits correspondingly stronger.  In Fig.~\ref{fig:sigmagg-fig2},
we show how our results would change if we had used an NFW profile
instead of the Kravtsov profile.  At most energies, the changes are
modest, and illustrative of the potential uncertainties.
The only significant change to the
combined gamma-ray limit is for the H.E.S.S. Galactic Center Ridge
case, which is based on small angular radii.
In the NFW case, the steeper profile gives an overall larger intensity
and a smaller signal cancelation when the background is subtracted.
A fuller discussion of how the annihilation signals depend on the choice
of dark matter density profile is given in YHBA.


\begin{figure}
\includegraphics[width=3.25in, clip=true]{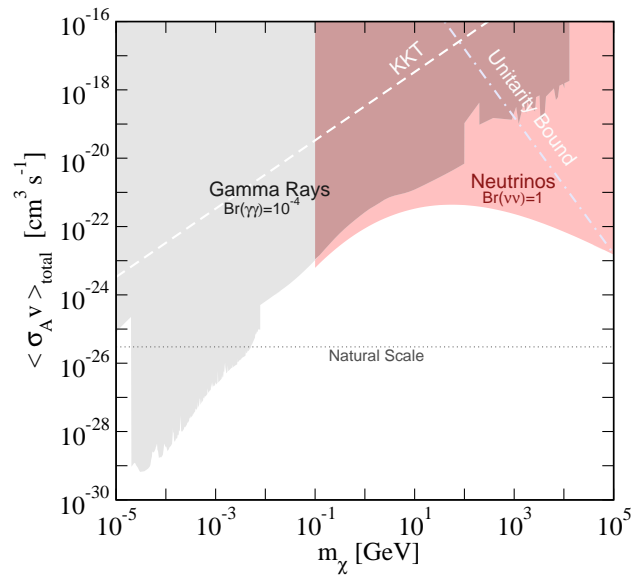}
\caption{
  The gamma-ray and neutrino limits on the total annihilation
  cross section, selecting $Br(\gamma\gamma) = 10^{-4}$ as a
  conservative value.  The unitarity and KKT bounds are also shown.
  The overall bound on the total cross section at a given mass is
  determined by the strongest of the various upper limits.}
  \label{fig:sigmagg-fig3}
\end{figure}

\subsection{Limits on the Total Cross Section}

Unsurprisingly, the cross section bounds derived under the
assumption of monoenergetic gamma rays are substantially
stronger than those defined similarly for final-state neutrinos in
BBM~\cite{Beacom:2006tt} and YHBA~\cite{Yuksel:2007ac}.
(At the highest masses, near $10^4$ GeV, this is no longer
true, first because of how the numerical limits work out, and then
because we do not presently have good gamma-ray data
or limits at higher energies; we expect that dedicated analyses
by H.E.S.S. and other experiments will soon improve this.)
Indeed, this was an assertion in those two works that we have
now justified in more detail than before.

It is unrealistic to have $Br(\gamma\gamma) = 100\%$, of course, if
one is trying to set a limit on the {\it total} cross section.  If
$Br(\gamma\gamma)$ is known, then a limit on the total cross section
can be determined by dividing the limit on the partial cross section
to that final state by the branching ratio:
\begin{equation}
\langle \sigma_A v \rangle_{total} =
\frac{\langle \sigma_A v \rangle_{\gamma\gamma}}{Br(\gamma\gamma)}.
\end{equation}
In typical models, this branching ratio is typically $10^{-3}$
or smaller~\cite{Jungman:1995df, Bertone:2004pz, Bergstrom:2000pn}.
To be conservative, we must just choose a value such that it is implausible
that the true branching ratio could be smaller.  We therefore assume
$Br(\gamma\gamma) = 10^{-4}$, but this choice could be debated.
As noted, our analysis uses wide logarithmic energy bins, and so, at the
very least, would capture the gamma rays near the endpoint due to 
internal bremsstrahlung from charged particles~\cite{RC}.
(Similarly, as a general point, limits on the total cross section defined
by assuming only $W^+ W^-$ final states~\cite{Hooper:2008zg}
would have to be corrected by dividing by $Br(W^+ W^-)$.)

Figure~\ref{fig:sigmagg-fig3} summarizes various limits on the total cross
section, including the one just described, the unitarity bound
mentioned earlier, and the neutrino bound from YHBA (based on the
Milky Way signal and the Kravtsov profile).  The standard cross section
for a thermal relic is also shown.
Note that our limits bound $\langle \sigma_A v \rangle$ directly,
independent of whether $\sigma_A$ is s-wave or p-wave dominated.
These results, combined with those in our Fig.~\ref{fig:sigmagg-fig2},
strongly constrain the possibilities for large dark matter annihilation
signals, e.g., as assumed in Ref.~\cite{Profumo:2008fy}.

When shown in this way, it becomes clear how surprisingly strong the
neutrino bound on the {\it total} cross section is, as it is
comparable to the bound obtained using the gamma-ray flux limits and
a reasonable assumption about the minimum branching ratio to gamma rays.
It is very important to emphasize
that while the gamma-ray bound on the partial cross section had to be
divided by a realistic $Br(\gamma\gamma)$, this is {\it not} the case
for the neutrino bound, as explained above.  If we assume only SM final
states, then all final states besides neutrinos lead to appreciable
fluxes of gamma rays, and hence are more strongly excluded.  Of
course, the gamma-ray and neutrino cross section limits can both be
weakened by assuming an appreciable branching ratio to new and truly
sterile particles.


\subsection{Conclusions and Prospects}

Using gamma-ray data from a variety of experiments, we have calculated
upper limits on the dark matter annihilation cross section to gamma rays over
a wide range of masses.   These limits are conservatively defined, in terms
of our analysis criteria, our assumptions about the uncertain dark matter
density profiles, and the gamma-ray spectrum.  While our results were
nominally defined for monoenergetic gamma rays with $E_\gamma = m_\chi$,
we have shown that all of our results except the INTEGRAL line flux limit
are only weakly dependent on this assumption.  The limits obtained for more
general gamma-ray spectra would only be somewhat less stringent.

There are good prospects for improved sensitivity with present and
upcoming gamma-ray experiments, particularly
GLAST~\cite{Morselli:2004ke,Morselli:2007fx, Baltz:2008wd} and the TeV ACT
detectors.  More detailed searches and analyses by the experimental
collaborations themselves should also lead to improvements, which we
encourage.  These searches for dark matter signals are already
background-limited, which will limit the possible improvements.  GLAST
and other experiments should be able to make reductions in the
backgrounds by taking advantage of better energy and angular
resolution, and by reducing the residual diffuse emission by
subtracting astrophysical components and resolving individual sources.
The high statistics expected for GLAST and other experiments should
also make it possible to define detection criteria in terms of the
uncertainty on the background, instead of the whole measured
background.

Using a conservative choice on the branching ratio to gamma rays,
namely $Br(\gamma \gamma) \simeq 10^{-4}$, we defined an upper limit
on the {\it total} dark matter annihilation cross section by dividing our
limits on the partial cross section to gamma rays by this branching
ratio.  At intermediate energies, the upper limit on the total cross
section defined this way is comparable to previous upper limits
defined using neutrinos~\cite{Beacom:2006tt, Yuksel:2007ac, 
PalomaresRuiz:2007eu}.  The
combined limit is considerably stronger than the unitarity
bound~\cite{Griest:1989wd, Hui:2001wy}, or
the cross section of Ref.~\cite{Kaplinghat:2000vt}, which would lead
to substantial modifications of dark matter halos.
For the relatively large cross sections considered here, the dark matter
could not be a thermal relic; additional work is needed to push the
sensitivity of these and other techniques down to the expected cross
section scale for thermal relics.


\medskip

{\bf Acknowledgments:}
We thank Brian Baughman, Matt Kistler, Shmuel Nussinov, Sergio Palomares-Ruiz,
and Casey Watson for useful discussions and comments.
GDM was supported by Department of Energy Grant DE-FG02-91ER40690,
TDJ by an Australian Postgraduate Award, NFB by the University of Melbourne
Early Career Researcher and Melbourne Research Grant Schemes, and
JFB and HY by NSF CAREER Grant PHY-0547102 to JFB.


\end{document}